\documentclass[sigconf,authordraft,review=false]{acmart}
\setcopyright{none}
\renewcommand\footnotetextcopyrightpermission[1]{} % removes footnote with conference information in first column
\usepackage{subcaption}
\usepackage{tikz}
\usepackage{makecell}
\tikzstyle{process} = [rectangle, rounded corners, minimum width=3cm, minimum height=1cm, text width=3cm, text centered, draw=darkblue, fill=white]
\tikzstyle{arrow} = [thick,->,>=stealth]
\usetikzlibrary{shapes,decorations,arrows,calc,arrows.meta,fit,positioning}
\tikzset{
    -Latex,auto,node distance =1 cm and 1 cm,semithick,
    state/.style ={ellipse, draw, minimum width = 0.7 cm},
    point/.style = {circle, draw, inner sep=0.04cm,fill,node contents={}},
    bidirected/.style={Latex-Latex,dashed},
    el/.style = {inner sep=2pt, align=left, sloped}
}

\DeclareMathOperator*{\argmax}{\arg\!\max}

\AtBeginDocument{%
  \providecommand\BibTeX{{%
    \normalfont B\kern-0.5em{\scshape i\kern-0.25em b}\kern-0.8em\TeX}}}

\setcopyright{acmcopyright}
\copyrightyear{2023}
\acmYear{2023}
\acmDOI{XXXXXXX.XXXXXXX}

\acmConference[]{
}{}{}

\acmPrice{}
\acmISBN{}
\begin{document}

\title{ACE: Active Learning for Causal Inference with Expensive Experiments}

\author{Difan Song}
\affiliation{%
  \institution{Georgia Institute of Technology}
  \city{Atlanta}
  \state{GA}
  \country{USA}
}
\email{dfsong@gatech.edu}

\author{Simon Mak}
\affiliation{%
  \institution{Duke Univeristy}
  \city{Durham}
  \state{NC}
  \country{USA}
}
\email{sm769@duke.edu}

\author{C.F. Jeff Wu}
\affiliation{%
  \institution{Georgia Institute of Technology}
  \city{Atlanta}
  \state{Georgia}
  \country{USA}
}
\email{jeff.wu@isye.gatech.edu}


\begin{abstract}
Experiments are the gold standard for causal inference. In many applications, experimental units can often be recruited or chosen sequentially, and the adaptive execution of such experiments may offer greatly improved inference of causal quantities over non-adaptive approaches, particularly when experiments are expensive. We thus propose a novel active learning method called ACE (Active learning for Causal inference with Expensive experiments), which leverages Gaussian process modeling of the conditional mean functions to guide an informed sequential design of costly experiments. In particular, we develop new acquisition functions for sequential design via the minimization of the posterior variance of a desired causal estimand. Our approach facilitates targeted learning of a variety of causal estimands, such as the average treatment effect (ATE), the average treatment effect on the treated (ATTE), and individualized treatment effects (ITE), and can be used for adaptive selection of an experimental unit and/or the applied treatment. We then demonstrate in a suite of numerical experiments the improved performance of ACE over baseline methods for estimating causal estimands given a limited number of experiments.

% More importantly, the GP model allows for uncertainty quantification of a variety of causal estimands, including the average treatment effect (ATE), the average treatment effect on the treated (ATTE), and individualized treatment effects (ITE).
\end{abstract}






\maketitle

\section{Introduction} \label{sec:intro}

In many real-world scenarios, we rely on experiments to gauge the impact of a particular action, policy, or change. Examples include clinical trials to test the effect of a new treatment, phased introduction of government policies before rolling out on a larger scale \cite{jowell2003}, and large-scale experiments that are one of a kind, such as the Tennessee STAR project that focused on issues in education \cite{mosteller1995tennessee}. Recent technological advancements gave rise to online experiments, including mobile health applications \cite{nahum2018just} and thousands of experiments conducted daily in internet companies \cite{kohavi2020trustworthy}. Such experiments often focus on the inference of \textit{causal estimands}, such as the average treatment effect (ATE) and average treatment effect on the treated (ATTE) on the population level or the individualized treatment effect (ITE) on the individual level.

% \textit{Randomization} and \textit{random sampling} are often regarded as two critical factors of a successful experiment. Randomization means that treatments are assigned randomly and independently for each experimental unit. If properly implemented, randomization maintains the balance of observable and unobservable attributes in different treatment groups \cite{rubin1974estimating}. Random sampling, on the other hand, requires that the sample under study is representative of the target population. Only then can the conclusions from the experiment generalize to broader scopes.

In many of the above applications, experiments can be performed (or are already conducted) in a sequential manner. Such adaptive experimentation has practical motivations in real-world scenarios. First and foremost, experiments are often highly \textit{expensive} (e.g., clinical trials), and it is thus desirable to maximize the efficiency with a limited experiment size.
% In other words, real-world experiments are seldom as simple as selecting a random sample, randomizing treatments, and observing the results.
Second, studies often span an extended period of time and must be performed sequentially as investigators work to increase precision (or reduce uncertainty) on causal estimands of interest. Moreover, dividing a study into several phases where the experiments ``ramp up'' to control the associated risks is common. In such cases, units sequentially enter the experiment, which makes acquiring information through active learning a natural choice. % Instead of strictly adhering to the randomization and random sampling principles, we pose the problem of whether we could gain from a sequential design of the experimental units.

Compared with experiments that emphasize random sampling and randomization of treatments, active learning is beneficial in several ways. First, active learning can target different quantities of interest (QoI). As we will discuss in Section \ref{sec:senarios}, different experiments have diverse objectives, and the cohorts of interest will also differ. Therefore, we can rely on active learning to acquire a representative sample that aligns with our target. Second, active learning makes use of learned structures to guide experiments. In other words, it \textit{exploits} the current knowledge about how units respond to the treatment to make informed decisions, thus more economical than a full-batch approach \cite{settles2011theories}. Finally, understanding the associated uncertainties in statistical models is crucial for increasing efficiency. With a small number of expensive experiments, we need to gradually \textit{explore} the decision space and acknowledge the uncertainties in the model. Only then can we generalize our model findings and confidently make decisions in an uncertain environment.

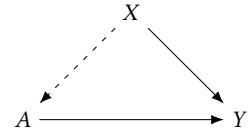
\begin{figure}[t]
    \centering
    \begin{tikzpicture}
    \node (1) at (0,0) {$X$};
    \node (2) [below left = of 1] {$A$};
    \node (3) [below right = of 1] {$Y$};

    \path[dash pattern=on 2pt off 3pt] (1) edge (2);
    \path (2) edge (3);
    \path (1) edge (3);
    \end{tikzpicture}
    \caption{Basic structure of the problems studied in this work. $X$ denotes the confounders, $A$ is the treatment, and $Y$ is the outcome of interest.}
    \Description{A diagram in which $X$ points to $A$ and $Y$, $A$ points to $Y$.}
    \label{fig:confounding}
\end{figure}

Throughout the rest of the paper, we work under the basic structure of the causal diagram in Figure \ref{fig:confounding}. $X$ represents the pre-experimental attributes, which we assume to be observable for all units in the pool before the experiment. $A$ is a dichotomous variable with $A_i = 1$ and $A_i = 0$ denoting that the $i$-th unit is in the treatment and control group, respectively. $Y$ is the outcome of interest, which consists of two potential outcomes $Y_i^{(1)} = Y_i(A_i = 1)$ and $Y_i^{(0)} = Y_i(A_i = 0)$ for each unit. We differentiate between two cases: (1) when the experimenter can decide $A$, the directed edge between $X$ and $A$ is removed; (2) when the experimenter can only decide $X$, the treatment $A$ is observed during the experiment. In case (2), the classical assumptions (individualistic, probabilistic, and unconfoundedness) for a regular assignment mechanism apply. See \cite{imbens2015causal} for details.

Using active learning in causal inference for problems with the above structure is an under-explored topic. \cite{sundin2019active} proposed a criterion based on the type-S error rate for estimating the ITE. \cite{jesson2021causal} also focused on the ITE and developed several acquisition functions based on Bayesian active learning by disagreement (BALD). However, both these works focused on sampling from \textit{observational data}. In particular, they assume that $(X, A)$ are observable, and we choose for which units we would like to observe $Y$. However, we find this setting to be less useful in practice because we seldom observe $A$ before the experiment. The treatment is either assigned as the unit enters the study or observed as the experiment progresses. Therefore, we always base the sequential designs on observed $X$ in this work.

The main contributions of this paper are two-fold: First, we identify several real-world applications in which a sequential design can benefit objectives in causal inference. We classify them into three scenarios where the available design components and goals differ. Naturally, our second contribution is to develop specific strategies for each scenario. In particular, we use Gaussian processes (GPs) to model the potential outcomes. Based on the uncertainty quantification provided by GPs, we develop acquisition functions that suit different objectives.

The rest of the paper is organized as follows. Section \ref{sec:senarios} extends the discussions in Section \ref{sec:intro} by describing three scenarios in more detail. Section \ref{sec:method} describes the Bayesian nonparametric modeling framework and our proposed strategy for the sequential design problem. Section \ref{sec:simulation} contains simulation studies to show the effectiveness of our method. Section \ref{sec:conclusion} concludes.

\section{Three scenarios for applications} \label{sec:senarios}

This section discusses motivating applications where sequential designs can be impactful. We classify them into three scenarios, and a summary is provided in Table \ref{tab:scenarios}.

\begin{table}[t]
    \centering
    \caption{Summary of the scenarios in Section \ref{sec:senarios}, with corresponding QoI).}
    \begin{tabular}{cccc} \toprule
    Scenario & Design & Observation & QoI \\ \midrule
    1 & $A_i$ & \makecell{$X_i$ (Before) \\ $Y_i$ (After)} & ATE \\
    2A & $X_i, A_i$ & $Y_i$ & ATE \\
    2B & $X_i$ & $A_i, Y_i$ (After) & ATE, ATTE, ... \\
    3 & $X_i$ & $A_i, Y_i$ (After) & ITE \\ \bottomrule
    \end{tabular}
    \label{tab:scenarios}
\end{table}

\subsection{Traditional random controlled trials}

By traditional random controlled trials (RCTs), we refer to the following setting: participants are recruited with consent; they understand when and where the experiment takes place; however, they may not know the exact treatment received (single-anonymous). Therefore, we cannot control which units enter the study, but we can assign the treatment $A_i$ according to the observed $X_i$. The goal in this Scenario 1 is thus to accurately estimate the ATE. A common characteristic of traditional RCTs is the high monetary and/or time costs required for experimentation. Thus, careful planning before the experiment and timely adjustments during the experiment are both critical. Here, we highlight below two instances of this in behavioral experiments and clinical trials.

 % In Section \ref{sec:intro}, we mentioned several applications that are examples of RCTs, showing their broad applicability.

Informally speaking, behavioral experiments aim to study how humans respond to certain stimuli, whether visual, audio, or specific instructions. These experiments are crucial in many fields, including psychology, cognitive science, economics, and marketing. The outcome of the experiments includes responses to questions or performance in tasks, but it can also be physical measurements such as hormone levels and brain activity. The experiments often contain the following steps: (1) setting up the environment for the experiment; (2) recruiting participants, often with monetary compensation; (3) assigning the treatment and conducting the experiment; (4) observing the outcomes. In practice, the remaining steps for a single participant only take a short time after setting up the experiment. Therefore, participants sequentially entering the experiment allows for sequential treatment assignment.

Moreover, it is possible to acquire background information during the recruitment step or at the beginning of the experiment. For instance, when using a software interface, the first section can ask questions or conduct tests that provide the information $X$. If the outcome of interest is the response time, we can assign different treatments to units with similar \textit{initial} response times. A mathematical formulation focusing on variance reduction of the ATE will be presented in Section \ref{sec:method}.

For clinical trials, the general procedure is similar to the previous four steps. However, in RCTs that test the effect of a new treatment or drug, discriminating between different patients will be unethical. For this reason, many clinical trials require double anonymity. Therefore, we must be more cautious and consider the specific problem. As an example, an interesting problem in disease prognostics is whether text messages help raise awareness and mitigate risk factors after recovery. In the study of \cite{chow2015effect}, the experiment lasted for years, while the typical follow-up time is three or six months after recovery. For this particular application, using information from earlier units is less prone to ethical concerns. Through this discussion, we mainly hope to emphasize that issues including ethics, privacy, and fairness arises in many applications, and we always need to address these issues before applying the strategies outlined in this work.

\subsection{Online experimentation}

In Scenario 2, we shift our focus to experiments in the online environment. A salient difference is that, while users consent when agreeing to the terms and conditions, they often become part of an experiment without even realizing it. In this case, companies can choose which users to include in the experiment rather than receiving them as inputs, which offers more flexibility in planning experimental campaigns.

\noindent\textbf{\textit{A. When the treatment can be assigned.}}

We first consider a setting similar to Scenario 1, where we are again interested in the ATE. During the experiment, on top of choosing which unit to include in the study, we can freely assign a treatment to the unit, i.e., we are designing the tuple $(X_i, A_i)$. An instance of this would be introducing a small feature change in the mobile application, potentially affecting all users. Experiments testing the impact on a small set of users have become standard practice, as even a minor change can unexpectedly influence user engagement, conversion, and other statistics. During implementation, the experiments go through \textit{ramping} where traffic gradually increases to control unknown risks \cite{kohavi2020trustworthy, xu2018sqr}.
%The ramping process balances speed, quality, and risk (SQR, \cite{xu2018sqr}) while gradually increasing the number of users exposed to the treatment.
Therefore, a sequential design naturally fits in with this procedure.

In the potential outcomes framework, we can consider the problem as estimating two response surfaces: one for the treatment group and one for the control group. Due to the possible effect heterogeneity, we must explore the entire feature space to estimate the ATE accurately. Thus, a sequential design favors unexplored regions when increasing the number of units. As the experiment size grows, we will likely have a design in the attributes $X$ that evenly fills the feature space for both response surfaces (see, e.g., existing work in ``space-filling'' designs \cite{joseph2016space}).

\noindent\textbf{\textit{B. When the treatment can only be observed.}}

\begin{table}[t]
    \centering
    \caption{Summary of different target populations, QoI, and associated weights.}
    \begin{tabular}{lcc} \toprule
    Target population & QoI & Weight $w(x)$ \\ \midrule
    Combined & ATE & $1$ \\
    Treated & ATTE & $e(x)$ \\
    Overlap & ATO & $e(x)(1 - e(x))$ \\
    Truncated combined & & $\mathbb{I}(\alpha < e(x) < 1 - \alpha)$ \\
    Matching & & $\min\{e(x), 1 - e(x)\}$ \\ \bottomrule 
    \end{tabular}
    \label{tab:QoI}
\end{table}

So far, we have focused on \textit{randomized experiments}. In this section, we enter the realm of \textit{observational studies} by assuming we no longer have control over the treatment $A_i$. Continuing the topic of the mobile application experiment, sometimes the change can be significant, for instance, a restyle of the user interface. In these cases, the users are often offered a choice between the old and new versions. Consequently, we no longer control to which groups the users belong. In the structural model in Figure \ref{fig:confounding}, we cannot ignore the directed edge from $X$ to $A$, indicating the existence of confounding.

In what follows, we assume that unconfoundedness holds given the user characteristics $X$. We can thus define the propensity score: \begin{align}
    e(x) = \mathrm{Pr}(A_i = 1 | X_i = x).
\end{align} Given this assumption, we can define multiple quantities-of-interest (QoIs) in the form of weight average treatment effects (WATE): \begin{align}
    \tau_w = \frac{\int_{\mathcal{X}} w(x) (\mu^{(1)}(x) - \mu^{(0)}(x)) dx}{\int_{\mathcal{X}} w(x) dx}, \label{eqn:WATE}
\end{align} where $\mu^{(1)}(x)$ and $\mu^{(0)}(x)$ are the expected values of $Y^{(1)}$ and $Y^{(0)}$ given $X = x$, respectively. The choice of the weights $w(x)$ depends on our population of interest. $w(x) = 1$ for all units means we view all the units equally, corresponding to the usual ATE; if we are more interested in the units that receive the treatment, we can take $w(x) = e(x)$, which gives us the average treatment effect on the treated (ATTE). Table \ref{tab:QoI} (adapted from \cite{li2018balancing}) summarizes such QoIs.

Clearly stating the target population will be significant for the sequential selection of $X$. A design will now favor units belonging to regions that are unexplored and also have a large weight $w(x)$. Intuitively, we will be trying to create a sample similar to the target population of interest; this intuition is verified in Section \ref{sec:simulation}.

\subsection{Marketing}

In both Scenario 1 and 2, we focused on quantities defined by a population. However, the average effect is not of primary interest in many experiments. Instead, finding individuals with high ITEs may be more beneficial since this allows us to invest more resources with precision. Marketing campaigns provide an instance with extensive applicability. For instance, supermarkets hope to offer discounts that ultimately increase the total purchase; banks advertise new products to those most likely to invest; internet advertising platforms show advertisements to raise conversion as high as possible.

These real-world applications all have structures similar to Scenario 2B. First, the companies initiating the experiments have information on the users, including demographic information (although there can be limitations on using this for algorithm development) and past engagement. Next is an immediate outcome, such as whether to accept an offer or click a link. Although this sign-post outcome is valuable, the experimenters are ultimately interested in metrics that provide long-term value. These components constitute our framework's $(X, A, Y)$ variables.

Therefore, we can use the same modeling framework, but the acquisition function for sequential designs will differ. Exploration is encouraged for learning about the entire response surface, while an optimization problem requires an exploration-exploitation tradeoff. Thus, we want to explore and find users that respond positively to the campaigns (exploration). Once we are confident of our findings, we should lean our resources toward these users to maximize the impact of the marketing effort (exploitation).

\section{ACE methodology} \label{sec:method}

This section describes the proposed ACE methodology for sequential designs in the aforementioned three scenarios. We first introduce a GP learning framework for modeling potential outcomes, then propose different novel acquisition functions that target learning for the desired QoIs in each scenario (see Table \ref{tab:scenarios}).

\subsection{Gaussian process regression}

We briefly introduce GP regression, which we use to model the relationship between the attributes $X$ and the potential outcomes. Specifically, we build separate GP models for the conditional mean functions for the treatment and control groups. This allows for different smoothness levels for the two response surfaces, which is crucial for modeling heterogeneity of the treatment effects.

Given $n$ observations with attributes $\boldsymbol{x}_{n} = [x_1, \dots, x_n]$ and treatment $a$, the corresponding outputs $\boldsymbol{y}_{n} = [y_1, \dots, y_n]$ take a multivariate normal prior: \begin{align}
     \boldsymbol{y}_{n} \sim \mathcal{N}\left(m_0(\boldsymbol{x}_{n}), \Sigma_0(\boldsymbol{x}_{n}, \boldsymbol{x}_{n}) + \eta^2 I_n \right), \label{eqn:GP_prior}
\end{align} where $m_0(\cdot)$ is the mean function, $\Sigma_0(\cdot, \cdot)$ is a covariance kernel, and $\eta^2$ is the variance of a normally-distributed random error. There are many possible specifications for $m_0(\cdot)$ and $\Sigma_0(\cdot, \cdot)$, which contain hyperparameters that we can estimate by maximum likelihood. See \cite{santner2018design} for a detailed discussion. At any new point $x$ with the same treatment $a$, the conditional posterior of the expected outcome is also Gaussian:
\begin{align} \label{eqn:GP_posterior}
\begin{split}
    &\mu^{(a)}(x) | \boldsymbol{y}_{n} \sim \mathcal{N}\left(m_n(x), \sigma_n^2(x)\right), \\
    &m_n(x) = m_0(x) + \Sigma_0\left(x, \boldsymbol{x}_{n}\right) \left(\Sigma_0\left(\boldsymbol{x}_{n}, \boldsymbol{x}_{n}\right) + \eta^2 I_n\right)^{-1} \left(\boldsymbol{y}_{n} - m_0\left(\boldsymbol{x}_{n}\right)\right), \\
    &\sigma_n^2(x) = \Sigma_0(x, x) - \Sigma_0\left(x_i, \boldsymbol{x}_{n}\right) \left(\Sigma_0\left(\boldsymbol{x}_{n}, \boldsymbol{x}_{n}\right) + \eta^2 I_n\right)^{-1} \Sigma_0\left(\boldsymbol{x}_{n}, x\right).
\end{split} 
\end{align} Therefore, we can provide a prediction for any potential user receiving treatment $a$, along with the uncertainty of the prediction.

By specifying two priors for $a = 1, 0$, respectively, we can get posteriors for $\mu^{(1)}(x)$ and $\mu^{(0)}(x)$ as long as we have data from both groups. Since the estimations of hyperparameters are done in isolation, the two response surfaces are allowed to have different landscapes. We should ensure sufficient samples for both groups to control the uncertainty in all scenarios.

\subsection{Adaptive design strategies}

Before specifying the proposed acquisition functions, we make some additional assumptions. For Scenarios 1 and 2, the quantities in \eqref{eqn:WATE} are defined for a fixed distribution. We assume having a test set $\boldsymbol{x}_{\text{test}} = [x_1, \dots, x_{n_{\text{test}}}]$ that is a representative sample of the population of interest. This assumption is reasonable if (1) we have a sample from a previous study or external source; (2) we can get compressed data from a large user pool using techniques outlined in \cite{joseph2022split, vakayil2022data}. Then, the QoI becomes a finite-sample version of \eqref{eqn:WATE}: \begin{align}
    \tau_w = \frac{\sum_{k = 1}^{n_{\text{test}}} w(x_k) (\mu^{(1)}(x_k) - \mu^{(0)}(x_k))}{\sum_{i = 1}^{n_{\text{test}}} w(x_k)}, \label{eqn:QoI}
\end{align} In vector notation, we write the potential outcomes and sample weights as: \begin{align}
    \boldsymbol{\mu}^{(a)} &= \left(\mu^{(a)}(x_1), \dots, \mu^{(a)}(x_{n_{\mathrm{test}}})\right), \quad a = 0, 1, \\
    \boldsymbol{w} &= \left(w(x_1), \dots, w(x_{n_{\mathrm{test}}})\right).
\end{align}

\noindent \textbf{\textit{ACE: Design $A_i$ only (Scenario 1)}}

In this scenario, we observe $X_i = x_i$ when a new unit enters the experiment. Given the dichotomous treatment variable, the experimenter can assign them to either the treatment or control group. We now consider the joint distribution of the expected potential outcomes $\boldsymbol{\mu}^{(a)}$ and $\mu^{(a)}(x_i)$. Augmenting \eqref{eqn:GP_posterior} with a superscript $(a)$ to indicate the treatment group, we can write the variance-covariance matrix of this distribution as: \begin{align}
\begin{bmatrix}
    \Sigma_n^{(a)}(\boldsymbol{x}_{n_{\text{test}}}, \boldsymbol{x}_{n_{\text{test}}}) & \Sigma_n^{(a)}(\boldsymbol{x}_{n_{\text{test}}}, x_i) \\
    \Sigma_n^{(a)}(x_i, \boldsymbol{x}_{n_{\text{test}}}) & \left(\sigma_n^{(a)}(x_i)\right)^2
\end{bmatrix}.
\end{align}

We can then use the conditional variance equation once more to get the posterior variance \textit{if we were able to observe $\mu^{(a)}(x_i)$}. Comparing the expression with the current posterior variance $\Sigma_n^{(a)}(\boldsymbol{x}_{n_{\text{test}}}, \boldsymbol{x}_{n_{\text{test}}})$, it is straightforward to calculate the \textit{variance reduction}: \begin{align} 
    r(x, a; \boldsymbol{w}) = \boldsymbol{w}^T \Sigma_n^{(a)}(\boldsymbol{x}_{n_{\text{test}}}, x) \left(\sigma_{n}^{(a)}\right)^{-2} \Sigma_n^{(a)}(x, \boldsymbol{x}_{n_{\text{test}}}) \boldsymbol{w}. \label{eqn:var_reduction}
\end{align} For Scenario 1, the weight $\boldsymbol{w}$ is a vector of 1's since the QoI is the ATE. We simply calculate this criterion for $x = x_i, a = 1, 0$ and select the treatment offering a larger variance reduction. A salient feature of GP regression is that uncertainty reduces quickly around observed data points. Thus, the variance reduction criterion will favor the potential outcome with larger uncertainty, which leads to a design that balances the treatment and control groups.

\noindent \textbf{\textit{ACE: Designing $(X_i, A_i)$ (Scenario 2A)}}

In the remaining scenarios, we are free to select the study participants sequentially. Therefore, we assume the existence of a user pool $\mathcal{X}_{\text{pool}} = \{x_i\}_{i = 1}^{n_{\text{pool}}}$. Due to practical constraints such as budget and risk management, only a fraction of the users in the candidate pool enters the experiment.

In Scenario 2A, the experimenter has two choices: the user to include and their assignment. We can use the variance reduction criterion \eqref{eqn:var_reduction}, but the optimization is now over \textit{both} $x$ and $a$: \begin{align}
    \argmax_{x \in \mathcal{X}_{\text{pool}}, a \in \{0, 1\}} r(x, a; \boldsymbol{1}).
\end{align} In each step, we need to calculate the variance reduction criterion $2 n_{\text{pool}} - n$ times, where $n$ is the current number of users already in the study. By choosing the unit and treatment that maximizes variance reduction, we can obtain designs that evenly fill the feature space for both response surfaces $a = 0, 1$.

\noindent\textbf{\textit{ACE-E: Designing $X_i$ only (Scenario 2B)}}

In this scenario, transitioning from an assigned treatment to an observed treatment creates a further complication. Although we can calculate the variance reductions in \eqref{eqn:var_reduction}, the actual effect on QoI estimation depends on the realization of $A_i$. Under the probabilistic assignment assumption \cite{imbens2015causal}, the realization is uncertain at the design stage. To tackle this problem, we define the ACE-E criterion using the \textit{expected variance reduction}: \begin{align}
    Er(x; \boldsymbol{w}) = e(x) r(x, 1; \boldsymbol{w}) + (1 - e(x)) r(x, 0; \boldsymbol{w}).
\end{align} By weighting with the propensity score, this term estimates the average variance reduction when we include a particular unit in our study. Here, since the propensity score differs for each unit, the QoI is no longer limited to the ATE. Instead, we can take any QoI from Table \ref{tab:QoI} and use its corresponding weights for $\boldsymbol{w}$.

When using the expected variance criterion, the propensity score is unknown. Therefore, we need to replace it with an estimated propensity score. Any estimation method based on the available data is valid, although we advise using robust estimation methods. The sequential design criterion for Scenario 2B is: \begin{align}
    \argmax_{x \in \mathcal{X}_{\text{pool}}} \widehat{Er}(x; \boldsymbol{w}) = \hat e(x) r(x, 1; \hat{\boldsymbol{w}}) + (1 - \hat e(x)) r(x, 0; \hat{\boldsymbol{w}}),
\end{align} where $\hat{\boldsymbol{w}}$ denotes the vector of weights with the propensity scores replaced with their estimated version.

\noindent \textbf{\textit{ACE-UCB: Designing $X_i$ for maximizing ITE (Scenario 3)}}

In Scenario 3, the setting is similar in that we have a pool of users as candidates. However, instead of focusing on a QoI, the objective becomes finding the units with a significant effect. In our notation, we would like to include $n$ users into the campaign that maximizes \textit{cumulative ITE} \begin{align}
    \sum_{i = 1}^n \mathbb{I}(A_i = 1) \left(\mu^{(1)}(x_i) - \mu^{(0)}(x_i)\right), \label{eqn:payoff}
\end{align} where $\mathbb{I}(\cdot)$ is an indicator function. This problem is more complicated than a classic optimization problem since it involves potential outcomes. Even with no noise, we can only observe one of the potential outcomes rather than the treatment effect. At any point of the experiment, our real-time evaluation of the total effect \eqref{eqn:payoff} almost always contains uncertainty.

Therefore, a sequential design should take uncertainty into account and address the exploration-exploitation tradeoff. We propose to use an upper-confidence bound (UCB) type acquisition function, inspired by the GP-UCB method \cite{srinivas2009gaussian}: \begin{align}
    \argmax_{x \in \mathcal{X}_{\text{pool}}} e(x) \left(\mu_n^{(1)}(x) - \mu_n^{(0)}(x)\right) + \beta_n^{1/2} \sigma_{\text{TE}}(x), \label{eqn:UCB}
\end{align} where the variance term is:
\begin{align} \label{eqn:var_TE}
\begin{split}
    \sigma_{\text{TE}}^2(x) &= e(x)\left(\left(\sigma_n^{(1)}\right)^2 + \left(\sigma_n^{(0)}\right)^2\right) \\
    &\quad + e(x)(1 - e(x)) \left(\mu_n^{(1)}(x) - \mu_n^{(0)}(x)\right)^2.
\end{split}
\end{align}
A detailed derivation of this can be found in the Appendix, and the specification of $\beta_n$ is discussed later in Section \ref{sec:simulation}. This criterion favors units with (1) large propensity scores, as only those who participate in the campaign are affected; (2) significant expected treatment effect for exploitation; (3) high uncertainty for exploration. Thus, this method is less likely to get stuck at local optima than greedy approaches. Similar to Scenario 2B, we also replace the propensity score with an estimator if necessary.
\section{Numerical experiments} \label{sec:simulation}

We now explore the performance of the proposed adaptive experimentation method in numerical experiments. For ease of visualization, we make use of the following two-dimensional set-up, modified from the popular Franke optimization test function \cite{franke1979critical}. With covariates $X \in [0, 1]^2$, the conditional mean functions take the form: \begin{align*}
    \mu^{(a)}(X = \boldsymbol{x}) &= \frac{3}{4}\exp\left(-\frac{1}{4}(9x_1 - 2)^2 - \frac{1}{4}(9x_2 - 2)^2\right) \\
    &\quad+ \frac{3}{4}\exp\left(-\frac{1}{49}(9x_1 + 1)^2 - \frac{1}{10}(9x_2 + 1)^2\right) \\
    &\quad + \frac{1}{2}a \exp\left(-\frac{1}{4}(9x_1 - 7)^2 - \frac{1}{4}(9x_2 - 3)^2\right) \\
    &\quad - \frac{1}{5}a\exp\left(-(9x_1 - 4)^2 - (9x_2 - 7)^2\right).
\end{align*}

\begin{figure}[tb]
    \centering
    \includegraphics[width=0.48\textwidth]{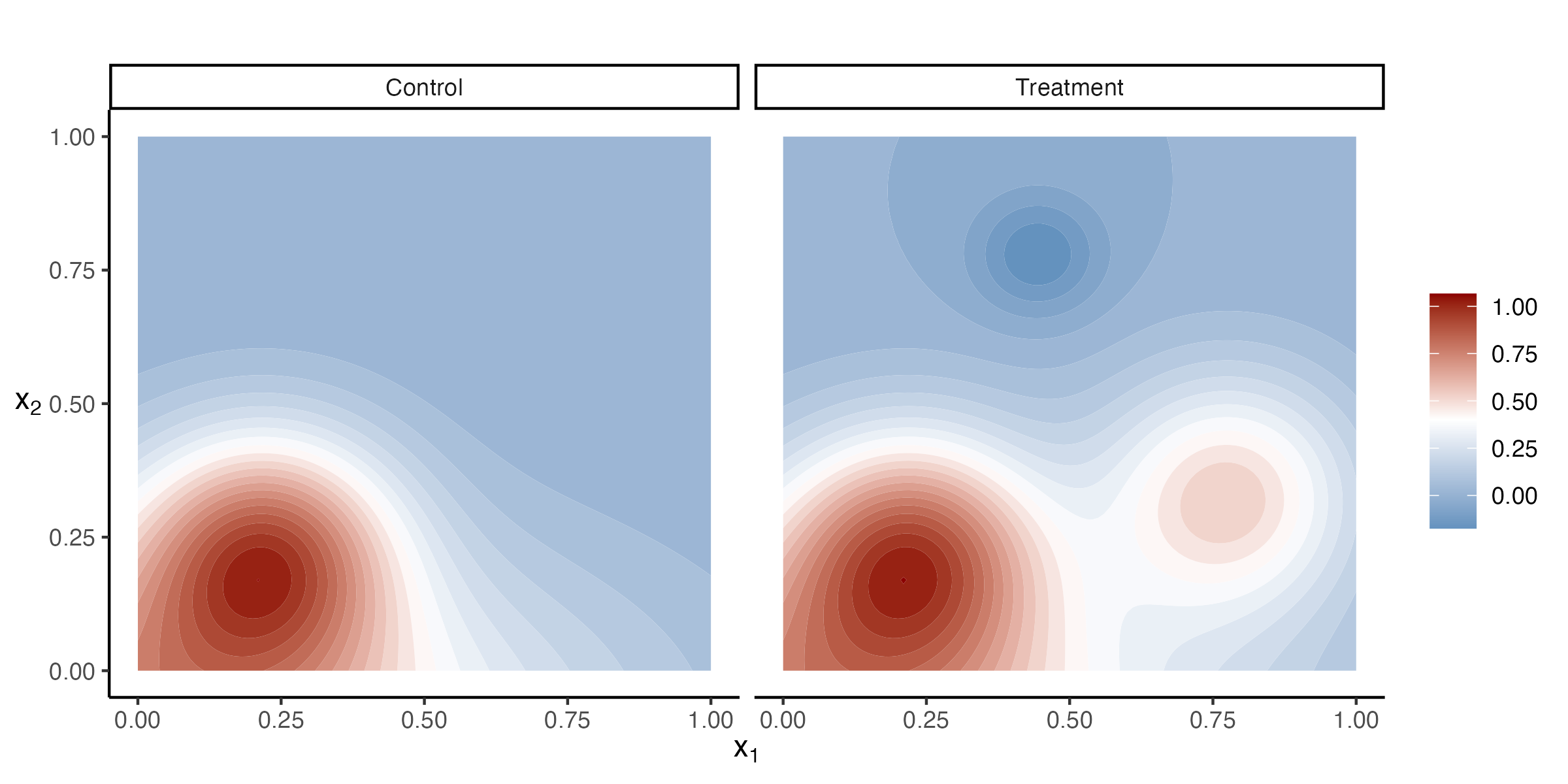}
    \caption{Potential outcomes for the treatment group and the control group.}
    \label{fig:Franke}
    \Description{Contour plots for target function when $a = 0$ (left) and $a = 1$ (right).}
\end{figure}
Figure \ref{fig:Franke} visualizes these functions for the treatment and control groups. We see multiple peaks in the function, which may represent cohorts of interest, e.g., high-value customers in banking applications. The treatment positively impacts the group with large $x_1$ and small $x_2$, while the group with large $x_2$ is negatively impacted.

\noindent \textbf{\textit{Scenarios 1 \& 2A.}}

\begin{table}[tb]
\centering
\caption{Scenarios 1 and 2A for ATE estimation. All results are enlarged by $10^3$.}
\label{tab:Sim1}
\begin{tabular}{lrrrr}
\toprule
& \multicolumn{2}{c}{Scenario 1} & \multicolumn{2}{c}{Scenario 2A} \\ \cmidrule(lr){2-3}\cmidrule(lr){4-5}
& Bias & RMSE & Bias & RMSE \\ \midrule
Random & 1.13 & 7.42 & 1.13 & 7.42 \\
ALC & $\boldsymbol{0.06}$ & $\boldsymbol{4.75}$ & 0.29 & 1.28 \\
ACE & -0.29 & 5.65 & $\boldsymbol{0.20}$ & $\boldsymbol{1.11}$ \\ \bottomrule
\end{tabular}
\end{table}

Table \ref{tab:Sim1} shows the results for ATE estimation in Scenarios 1 and 2A, where an experimenter can assign the treatment. Here, ``ALC'' is short for ``Active Learning Cohen'', which selects the point with the largest uncertainty $\sigma_n^{(a)}(x)$ (this is a widely-used GP active learning strategy; see \cite{seo2000gaussian,gramacy2020surrogates}). We obtain samples of size $n = 100$ to estimate the ATE. We report the bias and root mean squared error (RMSE) compared with the ground truth over 50 replications.

We can see both sequential methods outperform random sampling, which is not too surprising. In Scenario 1, ACE has no clear advantage over ALC, since the decision is dichotomous. However, when we introduce a pool of size $n_{\text{pool}} = 500$, ACE selects units that benefits the estimation the most, resulting in smaller RMSE.

\noindent \textbf{\textit{Scenario 2B.}}

% \begin{figure}[tb]
%     \centering
%     \includegraphics[width=0.30\textwidth]{Figures/Franke_propensity.png}
%     \caption{Propensity score with imbalance.}
%     \Description{Contour plot of the propensity score function.}
%     \label{fig:Franke_propensity}
% \end{figure}

For Scenarios 2B and 3, we use the propensity score function: \begin{align*}
    \mathrm{logit}(e(\boldsymbol{x})) = -2 + 2x_1 x_2,
\end{align*} where $e(\boldsymbol{x}) := \mathrm{Pr}(A = 1 | X = \boldsymbol{x})$. There is thus an imbalance, where the units in the treatment group are much fewer than in the control group (the overall propensity is around 20\%). In our simulations, we assume the propensity score is known for simplicity.

\begin{table}[tb]
\centering
\caption{Scenario 2B for estimating different QoI. All results are enlarged by $10^3$.}
\label{tab:Sim2}
\begin{tabular}{lrrrrrr} \toprule
& \multicolumn{2}{c}{ATE} & \multicolumn{2}{c}{ATTE} & \multicolumn{2}{c}{ATO} \\ \cmidrule(lr){2-3}\cmidrule(lr){4-5}\cmidrule(lr){6-7}
& Bias & RMSE & Bias & RMSE & Bias & RMSE \\ \midrule
Random & 5.56 & 45.80 & 2.62 & 12.83 & 5.70 & 13.68 \\
ALC-E & -13.55 & 49.92 & 0.69 & 4.56 & 1.93 & 5.49 \\
ACE-E & $\boldsymbol{3.55}$ & $\boldsymbol{32.92}$ & $\boldsymbol{0.45}$ & $\boldsymbol{2.88}$ & $\boldsymbol{0.05}$ & $\boldsymbol{3.14}$ \\ \bottomrule
\end{tabular}
\end{table}

We take $n = 100, n_{\text{pool}} = 500$ for the ATE and $n = 200, n_{\text{pool}} = 1,000$ for the ATTE and ATO due to the imbalance in the data. We compare the proposed ACE-E with the expected ALC (ALC-E) approach as baseline, defined as: \begin{align}
    \argmax_{x \in \mathcal{X}_{\text{pool}}} e(x) \left(\sigma_n^{(1)}(x)\right)^2 + (1 - e(x)) \left(\sigma_n^{(0)}(x)\right)^2.
\end{align}
This extends the existing ALC approach \cite{seo2000gaussian} to the scenario where treatments can only be observed.

\begin{figure}[tb]
  \centering
  \includegraphics[width=0.48\textwidth]{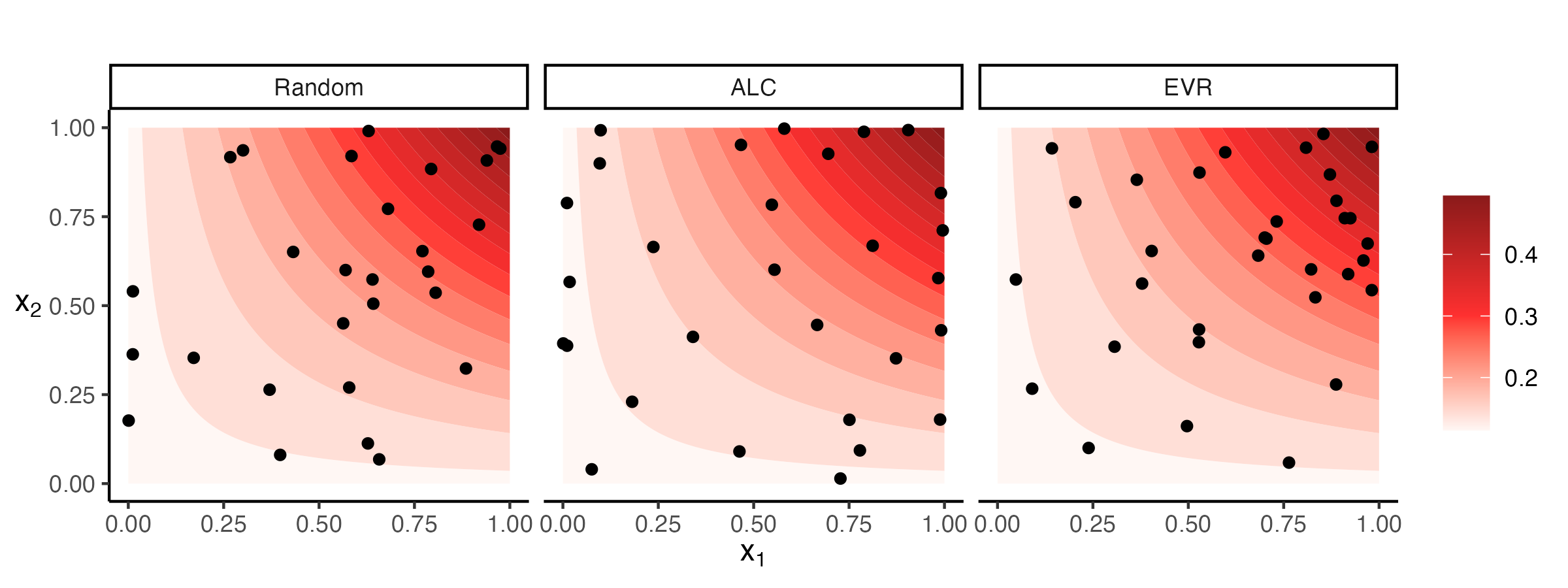}
  \caption{Visualization of the selected points in the treatment group for ATTE estimation using the three methods.}
  \Description{Three subfigures plotting each method's points in the treatment group. The background is the contour of the propensity score.}
  \label{fig:Franke_points}
\end{figure}
Results are shown in Table \ref{tab:Sim2}, where we observe that the proposed ACE-E has the smallest error in all cases. In Figure \ref{fig:Franke_points}, we show an example of the points selected by each method when the QoI is the ATTE. We plot the points in the \textit{treated group} since the small number of treated units confines the estimation accuracy. The points are plotted against contours of propensity scores, with darker colors indicating higher propensity. We can see that ALC-E is space-filling and selects many points near the boundary. However, each unit is weighted proportional to the propensity score in ATTE estimation. Our sequential design strategy based on ACE-E considers this weight in the acquisition. The resulting sample reflects this proportional relationship, which could explain why our method outperforms.

\noindent \textbf{\textit{Scenario 3.}}

\begin{figure}
    \centering
    \includegraphics[width=0.30\textwidth]{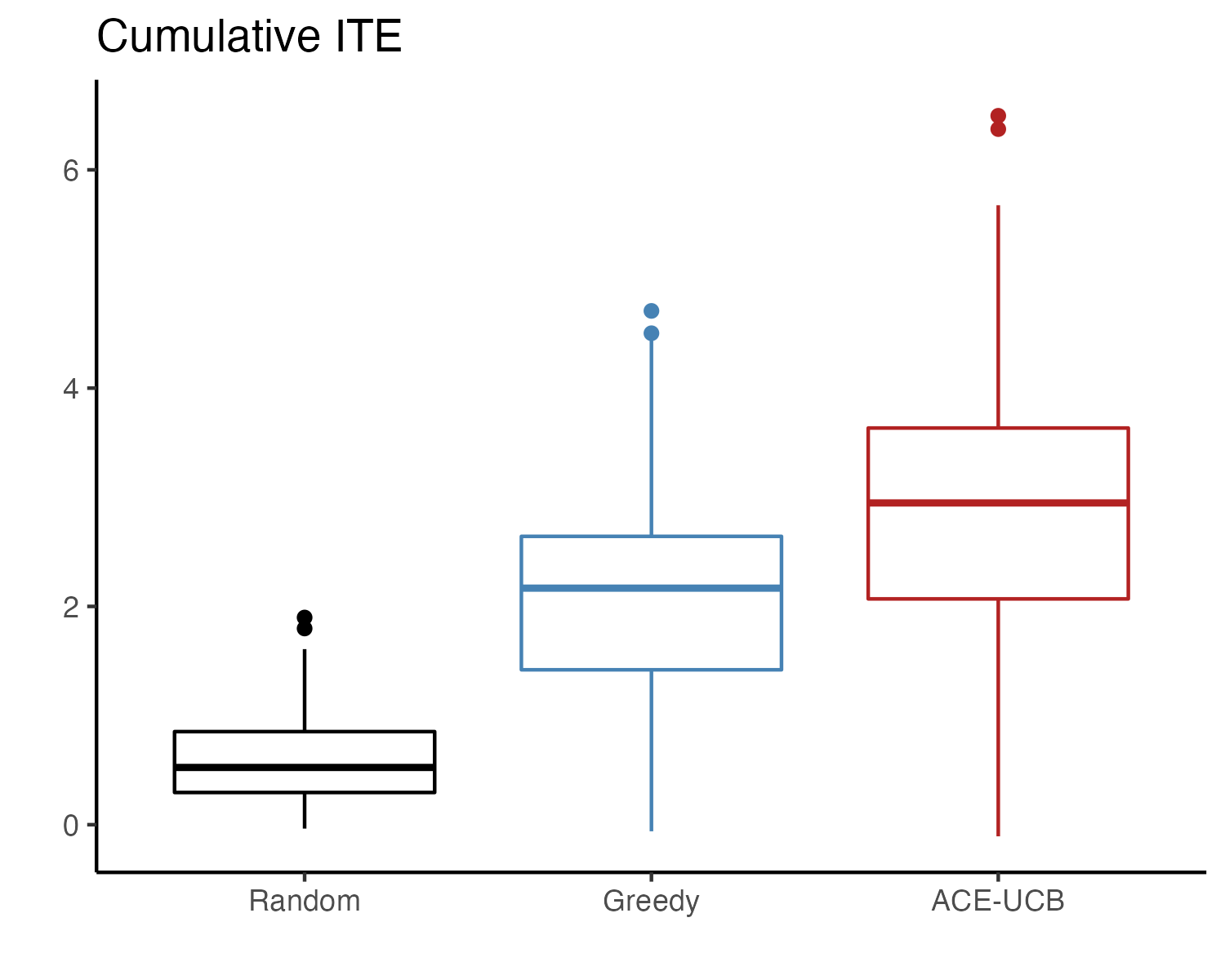}
    \caption{Cumulative ITE [Equation \eqref{eqn:payoff}] for different methods in Scenario 3.}
    \Description{Boxplots corresponding to three methods. The UCB is higher than greedy, which is higher than random.}
    \label{fig:Sim3}
\end{figure}

In Scenario 3, we compare the proposed ACE-UCB approach with the following baseline greedy approach: \begin{align}
    \argmax_{x \in \mathcal{X}_{\text{pool}}} e(x) \left(\mu_n^{(1)}(x) - \mu_n^{(0)}(x)\right).
\end{align} This acquisition exhibits full exploitation by directly maximizing the expectation of \eqref{eqn:payoff}. In comparison, \eqref{eqn:UCB} has an additional term encouraging exploration. In our method, we use $\beta_t = c^2 \log t$ (common choice in the UCB literature), with $c$ set as $0.01$. We take $n = 50, n_{\text{pool}} = 1,000$, and present the results of 50 replications in Figure \ref{fig:Sim3}. The boxplots show the superiority of the proposed ACE-UCB approach, which shows that it can effectively leverage the uncertainty from the adopted Bayesian model for targeted optimization of the ITE.

\section{Conclusion} \label{sec:conclusion}

We proposed in this work a new active learning method called ACE, which makes use of an underlying Gaussian process model of the conditional mean functions for guiding the adaptive selection of expensive experiments. ACE features a range of novel acquisition functions, which can target the estimation of a variety of causal estimands (e.g., ATE, ATTE, ITE) for three broad scenarios encountered in practical causal applications. We then showed the improved performance of ACE over several baseline methods with limited experiments in a suite of numerical experiments.

% Specifically, the uncertainty quantification properties of GPs allowed us to design acquisition functions based on the objectives.

% We surveyed a range of real-world scenarios where we had control over different factors, and the sequential nature of the experiments allowed for sequential planning of the factors.

While the proposed ACE approach appears promising, we are currently embarking on many avenues for future work. First, we are exploring batch-sequential modifications of the ACE acquisition functions, to reflect the fact that in practice, one would often conduct adaptive experiments in batches. Second, we are investigating the effectiveness of ACE in a variety of practical causal applications, including behavioral experiments conducted in laboratory environments \cite{data/VYCCZI_2017} and a real-world marketing campaign \cite{dunnhumby2014}. 

\begin{acks}
The authors thank Prof. Jared Huling from the University of Minnesota and Dr. Yuan Wang from Wells Fargo for the insightful discussions and helpful advice.
\end{acks}

\bibliographystyle{ACM-Reference-Format}
\bibliography{bibliography}


\appendix

\section{Derivation of the UCB variance}

Here we provide a short derivation of the variance term in Equation \eqref{eqn:var_TE}.

First, we have three random variables that are assumed to be independent given $X = x$: \begin{align*}
    \mathbb{I}(A = 1) &\sim Bernoulli(e(x)), \\
    \mu^{(1)}(x) &\sim \mathcal{N}\left(\mu_n^{(1)}(x), \left(\sigma_n^{(1)}\right)^2\right), \\
    \mu^{(0)}(x) &\sim \mathcal{N}\left(\mu_n^{(0)}(x), \left(\sigma_n^{(0)}\right)^2\right),
\end{align*}

From the relationship $\mathrm{Var}(XY) = \mathbb{E}(X^2) \mathrm{Var}(Y) + \mathrm{Var}(X)\mathbb{E}^2(Y)$, we can get the variance of $\mathbb{I}(A = 1)\left(\mu^{(1)}(x) - \mu^{(0)}(x)\right)$: \begin{align*}
    &\mathrm{Var}\left[\mathbb{I}(A = 1)\left(\mu^{(1)}(x) - \mu^{(0)}(x)\right)\right] \\
    =\ & \mathbb{E}(\mathbb{I}(A = 1)) \mathrm{Var}\left(\mu^{(1)}(x) - \mu^{(0)}(x)\right) \\
    &+ \mathrm{Var}(\mathbb{I}(A = 1))\mathbb{E}^2 \left(\mu^{(1)}(x) - \mu^{(0)}(x)\right) \\
    =\ &e(x)\left(\left(\sigma_n^{(1)}\right)^2 + \left(\sigma_n^{(0)}\right)^2\right) \\
    &+ e(x)(1 - e(x)) \left(\mu_n^{(1)}(x) - \mu_n^{(0)}(x)\right)^2.
\end{align*}

\section{Online resources}

The R codes used for the numerical example can be found at \url{https://github.com/difan1996/Causal-design}.

\end{document}